\documentclass[sn-aps]{sn-jnl}

\usepackage{graphicx}%
\usepackage{multirow}%
\usepackage{amsmath,amssymb,amsfonts}%
\usepackage{amsthm}%
\usepackage{mathrsfs}%
\usepackage[title]{appendix}%
\usepackage{xcolor}%
\usepackage{textcomp}%
\usepackage{manyfoot}%
\usepackage{booktabs}%
\usepackage{algorithm}%
\usepackage{algorithmicx}%
\usepackage{algpseudocode}%
\usepackage{listings}%
\usepackage{orcidlink}

\newcommand{\paper}[1][review]{#1}

\begin{document}

\title[Ultrahigh energy cosmic rays and neutrino flux models]{Ultrahigh energy cosmic rays and neutrino flux models}


\author[1,2,3]{\fnm{Marco Stein} \sur{Muzio}~\orcidlink{0000-0003-4615-5529}}\email{msm6428@psu.edu}

\affil[1]{\orgdiv{Department of Astronomy and Astrophysics}, \orgname{Pennsylvania State University}, \orgaddress{\city{University Park}, \postcode{16802}, \state{PA}, \country{USA}}}

\affil[2]{\orgdiv{Department of Physics}, \orgname{Pennsylvania State University}, \orgaddress{\city{University Park}, \postcode{16802}, \state{PA}, \country{USA}}}

\affil[3]{\orgdiv{Institute of Gravitation and the Cosmos, Center for Multi-Messenger Astrophysics}, \orgname{Pennsylvania State University}, \orgaddress{\city{University Park}, \postcode{16802}, \state{PA}, \country{USA}}}


\abstract{In this review we motivate ultrahigh energy neutrino searches and their connection to ultrahigh energy cosmic rays. We give an overview of neutrino production mechanisms and their potential sources. Several model-independent benchmarks of the ultrahigh energy neutrino flux are discussed. Finally, a brief discussion of approaches for model-dependent neutrino flux predictions are given, highlighting a few examples from the literature.}

\keywords{Ultrahigh energy neutrinos, ultrahigh energy cosmic rays, cosmic ray sources, neutrino astronomy, multimessenger astrophysics}



\maketitle

\section{Introduction}\label{sec:intro}

\par
The era of extragalactic neutrino astronomy was ushered in over a decade ago by the IceCube observatory upon detection of an astrophysical neutrino flux~\citep{IceCube:2013low}. This neutrino flux has now been robustly observed from ${\sim}10$~TeV to ${\sim}1$~PeV~\citep{Naab:2023xcz,Silva:2023wol} and a first point source, NGC 1068, has been identified with $4.2\sigma$ significance~\citep{IceCube:2022der}. IceCube has achieved myriad of other remarkable discoveries including: coincident detection of an astrophysical neutrino with flaring blazar TXS-0506+056~\citep{IceCube:2018cha,IceCube:2018dnn}, detection of a flux of neutrinos from the Galactic plane~\citep{IceCube:2023ame}, as well as, robust detections of a Glashow resonance event~\citep{IceCube:2021rpz} and tau neutrinos~\citep{IceCube:2024nhk}.

\par
However, ultrahigh energy (UHE) neutrinos ($E_\nu \gtrsim 10^{16.5}$~eV $\simeq 30$~PeV) have yet to be discovered. Observation of these particles would give unprecedented access to extreme astrophysical environments in the universe on cosmological scales. While the universe is opaque at high energies to other particle messengers, in particular cosmic rays (CRs) and photons, UHE neutrinos can propagate cosmological distances effectively unattenuated. Additionally, UHE neutrinos are undeflected by the Galactic and extragalactic magnetic fields, since they do not carry electric charge, meaning that their arrival directions can be used to straightforwardly determine their sources. Conditions inside their sources can also be probed by their flavor content at Earth. 

\par
UHE neutrinos would further provide a unique window into particle physics beyond LHC energies. Their observation would allow the neutrino-nucleon cross section to be measured at unprecedented center-of-mass energies, testing a number of beyond the Standard Model (BSM) theories. What's more, otherwise inaccessible BSM processes could potentially be probed by UHE neutrino observations, including additional neutrino interactions, sterile neutrinos, dark matter, and Lorentz invariance violation~\citep{Ackermann:2022rqc}.

\par
Despite the enormous astrophysics and particle physics potential of UHE neutrinos, an obvious question remains: do UHE neutrinos exist in Nature? And, if they do, what is their flux? While today these questions remain open, this \paper{} aims to provide an overview of their current understanding in the field. In Section~\ref{sec:background} we discuss the necessary conditions for UHE neutrinos to be produced and discuss their potential sources. In Section~\ref{sec:benchmarks} we overview some model-independent benchmarks of the UHE neutrino flux, while in Section~\ref{sec:predictions} we provide a brief overview of model-dependent predictions of their flux. Finally, in Section~\ref{sec:summary} we provide some concluding remarks and an outlook for the field over the coming decade.

\section{Background}\label{sec:background}

\subsection{Production channels}

\par
To understand whether UHE neutrinos may exist it is useful to first explore how they might be produced. For the purposes of this \paper{} we will focus on the production mechanisms within the Standard Model.

\par
Let us begin by investigating the possible production channels with the goal of determining the lowest energy stable particle which could produce a UHE neutrino. Neutrinos are produced in weak decays such as beta decay or charged pion decay. Let's consider each of these decay channels. 

\par
When a neutron undergoes a beta decay it produces a daughter proton, electron, and electron antineutrino,
\begin{align}
    n &\rightarrow p + e^- + \overline{\nu}_e~.
\end{align}
\noindent 
Since the neutrino is essentially massless, in the rest frame of the neutron it will have an energy of roughly $E_\nu^\mathrm{CM} \simeq m_n - m_p - m_e$. Boosting into the lab frame this implies that $E_\nu \simeq \frac{1}{1000} E_n$. 

\par
On the other hand, charged pion decay results in three neutrinos in the final state. First, the pion (which we take to be positive for illustration) decays into an antimuon and muon neutrino. The antimuon, itself unstable, will decay in a muon antineutrino, positron, and electron neutrino. The full decay chain can be written as,
\begin{align}
    \pi^+ &\rightarrow \nu_\mu + \mu^+ \nonumber \\
    &\rightarrow \nu_\mu + \overline{\nu}_\mu + \nu_e + e^+~.
\end{align}
\noindent
The first neutrino, from the initial pion decay, will carry an energy of roughly $E_\nu^\mathrm{CM} = \frac{m_{\pi^+}^2 - m_\mu^2}{2m_{\pi^+}}$ in the pion rest frame. Boosting to the lab frame this becomes $E_\nu \simeq \frac{1}{4} E_\pi$. The final two neutrinos, produced by the muon decay, will roughly share the muon energy equally between themselves and the electron, since $m_\nu, m_e \ll m_\mu$. Since the approximate muon energy is given by $E_\mu \simeq \frac{3}{4} E_\pi$, this implies that the final two neutrinos also have roughly $E_\nu \simeq \frac{1}{4} E_\pi$. 

\par
In summary, ignoring the sign of the lepton charge, beta decay produces a single electron neutrino with energy $E_\nu \simeq \frac{1}{1000} E_n$, while charged pion decay results in two muon neutrinos and one electron neutrino each with an energy $E_\nu \simeq \frac{1}{4} E_\pi$. Since we are interested in determining the lowest energy stable particle which could produce a UHE neutrino, we will set the neutron production channel aside and consider only pion production channels. Pions can be produced both photohadronically (i.e. via a nucleon-photon interaction) and hadronically (i.e. via a nucleon-nucleon interaction).

\par
 At production threshold, photohadronic interactions produce pions through a $\Delta$ resonance. This resonance primarily decays to a nucleon and pion. The full interaction be written as,
\begin{align}
    p+\gamma \rightarrow \Delta^+ \rightarrow p/n + \pi^0/\pi^+~.
\end{align}
\noindent
Following the same line of argument as above, in the $\Delta$ resonance rest frame the pion will have energy $E_\pi^\mathrm{CM} = \frac{m_\Delta^2 - m_{p/n}^2 + m_\pi^2}{2m_\Delta}$, which is roughly $E_{\pi^{0/+}} \simeq \frac{1}{5} E_\Delta$ in the lab frame. If the primary proton is such that $E_p \gg E_\gamma$, then simply $E_\pi \simeq \frac{1}{5} E_p$.

\par
On the other hand, hadronic interactions are generally much more complicated, in particular at high energies. In general one expects that a nucleon-nucleon interaction will produce a large number of particles, including many pions. Since the multiplicity of particles in the final state is large, the fraction of the primary nucleon energy carried by each pion will be much smaller than in the photohadronic production channel. For this reason, we will focus on the photohadronic channel.

\par
Therefore, the lowest energy stable particle which could produce a UHE neutrino would be a proton of energy $E_p = 20 E_\nu^\mathrm{UHE} \simeq 20 \times 10^{16.5}$~eV$\simeq10^{17.8}$~eV undergoing photopion production. This means that UHE neutrino production requires nucleons in the EeV energy regime: ultrahigh energy cosmic rays. 

\subsection{Constraints \& uncertainties from UHECR data} \label{sec:UHECRdata}

\par
\begin{figure}[htpb!]
    \centering
    \includegraphics[width=0.495\linewidth]{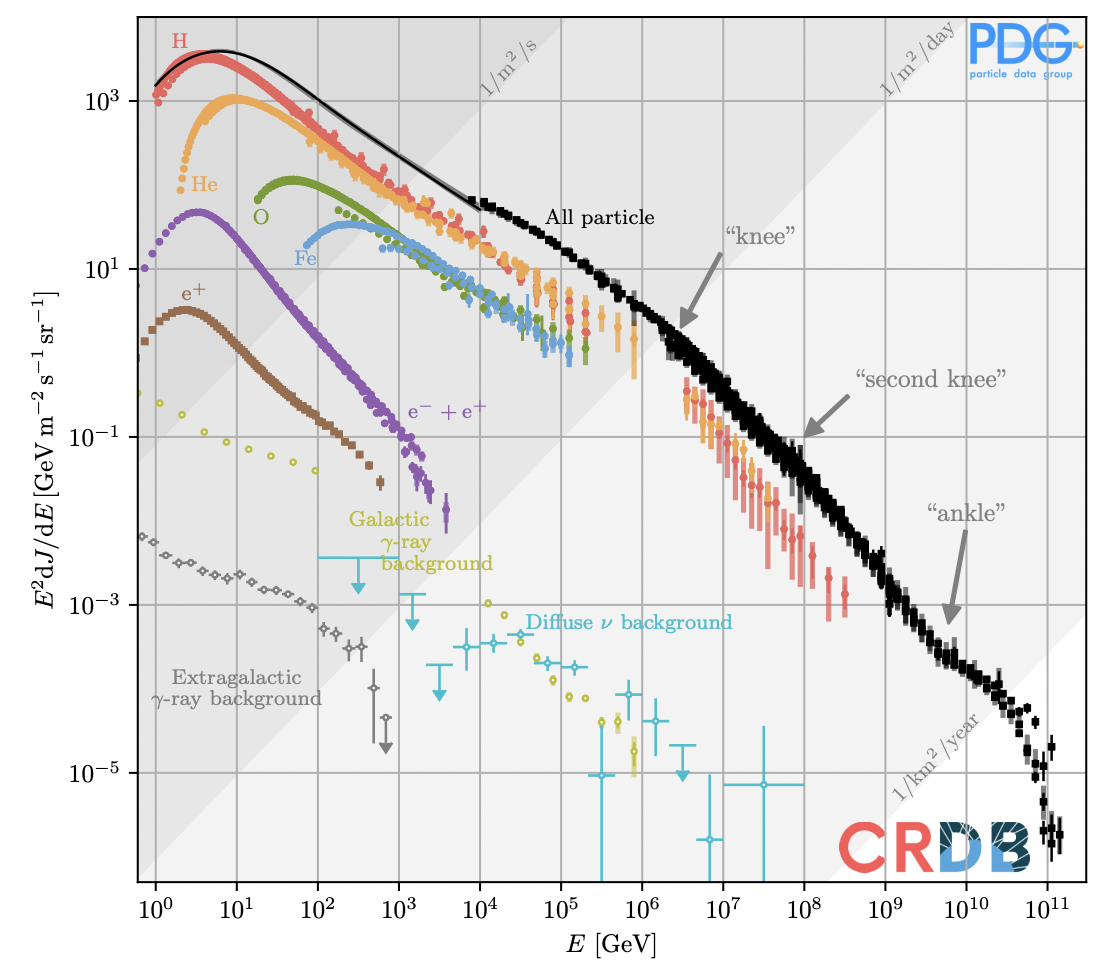}
    \includegraphics[width=0.495\linewidth]{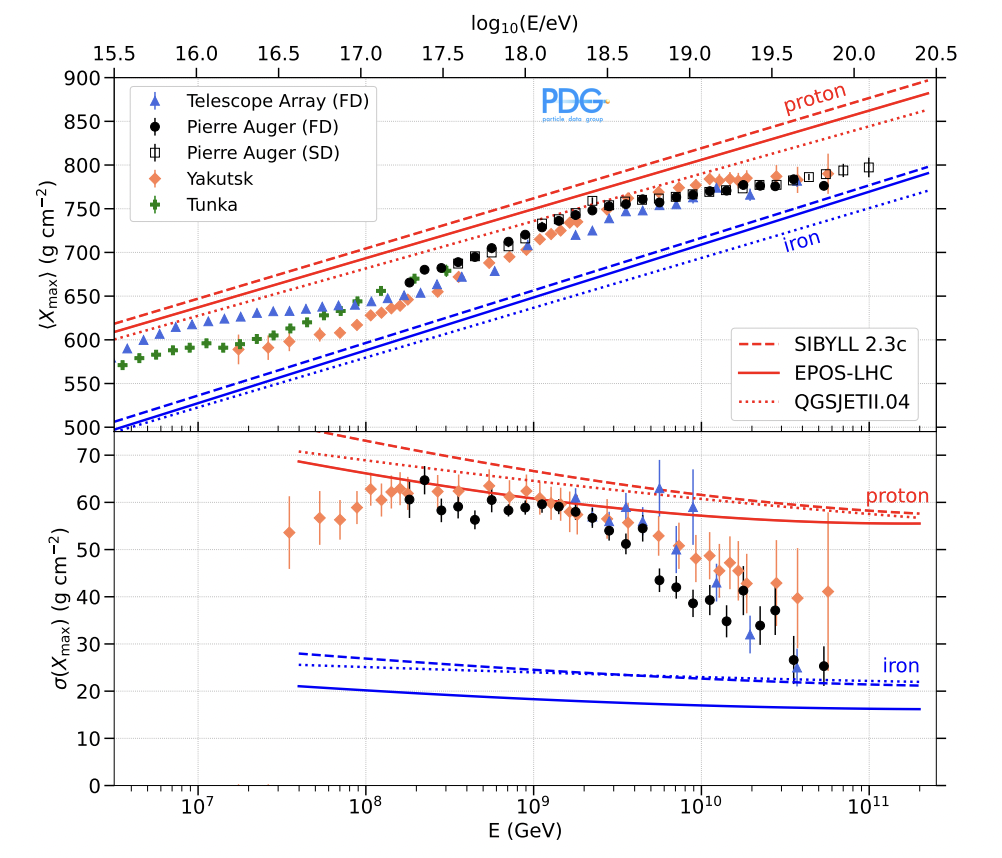}
    \caption{Summary of current UHECR spectrum (left) and composition (right) data. Reproduced from~\cite{pdg24}.}
    \label{fig:uhecrOverview}
\end{figure}

Ultrahigh energy cosmic rays (UHECRs) were first observed by the Volcano Ranch experiment in the 1960s~\citep{Linsley:1963km}, but today their sources remain unknown. Importantly, though, their existence demonstrates that Nature provides particles which could theoretically produce UHE neutrinos. This fact means that UHECRs are inextricably linked to UHE neutrinos and their data provides an important touchstone in modeling the UHE neutrino flux. In this section, we review only the most relevant facts about UHECRs; for a thorough review of the state of UHECRs see~\cite{Anchordoqui:2018qom,pdg24}. An overview of current UHECR spectrum \& composition data is shown in Fig.~\ref{fig:uhecrOverview}.

\par
The UHECR spectrum has been robustly measured up to ${\sim}10^{20.3}$~eV by the Pierre Auger Observatory and the Telescope Array~\citep{PierreAuger:2023wti}. Events at higher energies are suppressed due to a cutoff in the spectrum, observed to begin around $10^{19.7}$~eV. The origin of this cutoff is still an open question but there are two leading hypotheses: 1) that astrophysical sources are simply unable to accelerate CRs to higher energies -- sometimes referred to as sources ``running out of steam"; or 2) that the spectrum is suppressed due to UHECR interactions with the cosmic microwave background (CMB), often called the Greisen-Zatsepin-Kuzmin (GZK)  effect~\citep{Greisen:1966jv,Zatsepin:1966jv}. In the latter case, the spectrum is suppressed due to the large energy losses to photopion production from interactions of UHE protons with the CMB. Above $10^{19.7}$~eV the proton energy loss length due to these interactions is $\lesssim100$ Mpc, meaning the GZK effect induces a horizon on this scale~\citep{Anchordoqui:2018qom}. This mechanism is particularly relevant for the study of UHE neutrinos since such a process necessarily produces a sizable flux of EeV-scale neutrinos. 

\par
However, this prediction for a large flux of UHE neutrinos via the GZK effect is strongly dependent on the composition of CRs. Nuclei interacting with CMB photons will primarily photodisintegrate, rather than photopion produce, since their cross section is dominated by the giant dipole resonance~\citep{Anchordoqui:2018qom}. Since this interaction also causes substantial energy losses on similar length scales, it can also result in a suppression of the UHECR flux at similar energies. Therefore, the expected UHE neutrino flux due to nuclei above $10^{19.7}$~eV will be substantially suppressed compared to that produced by protons at these energies, even if the cutoff is due to the GZK effect for nuclei. For this reason, measurements of the UHECR (chemical) composition are needed in order to accurately assess the expected flux of UHE neutrinos from the GZK effect.

\par
Due to their low flux at Earth, $\lesssim 1\text{ CR}/\text{km}^2/\text{century}$, UHECRs cannot be measured directly. Instead, UHECRs are observed through the extensive air showers which they initiate in the atmosphere. The leading UHECR observatories today, the Pierre Auger Observatory and the Telescope Array, measure extensive air showers with two separate detectors: a ground detector array, which measures the shower as it reaches the ground, and fluorescence telescopes, which image the air shower development through the fluorescent light emitted along the shower axis. The intensity of the fluorescence is proportional to the number of particles in the shower, meaning that its observation provides a measure of the number of particles in the shower at a given time~\citep{Anchordoqui:2018qom}. In particular, this allows modern observatories to directly measure the depth of shower maximum, $X_\mathrm{max}$, via their fluorescence telescopes.\footnote{Recently deep learning techniques have enabled the extraction of $X_\mathrm{max}$ from ground detector measurements alone~\citep{PierreAuger:2024flk,PierreAuger:2024nzw}, significantly increasing the number of events with $X_\mathrm{max}$ data.} The depth of shower maximum is determined by both the primary particle energy and its mass number (really its energy-per-nucleon), but the exact mapping between these observables and $X_\mathrm{max}$ is unknown due to uncertainties in particle physics. However, a number of models for hadronic interactions exist (e.g.~\cite{Ostapchenko:2010vb,Pierog:2013ria,Riehn:2019jet}), allowing $X_\mathrm{max}$ data to be interpreted in terms of the primary particle's mass number after adopting a particular model.

\par
From the latest composition measurements from the Pierre Auger Observatory and Telescope Array~\citep{PierreAuger:2023yym} a clear picture has emerged: the composition of UHECRs at UHEs is not purely protonic. While the exact composition is still subject to sizable theoretical \& statistical uncertainties, it is evident qualitatively that the composition is getting heavier and more pure above $10^{18.6}$~eV. In particular, if protons exist above $10^{19.7}$~eV their abundance must be $<10\%$ of the total flux~\citep{Muzio:2019leu,Muzio:2023skc}, based on fluorescence detector measurements. This fact places significant constraints on UHECR models, which must explain both the UHECR spectrum and composition data, limiting the UHE neutrino flux which can be produced locally by the GZK effect. 

\par
However, UHECRs themselves only probe sources within $\lesssim 100$~Mpc of Earth, due to the horizon imprinted by the GZK effect (for both protons and nuclei). On the other hand, UHE neutrinos produced by UHECR interactions with the CMB and extragalactic background light (EBL) in propagation, often called \textit{cosmogenic neutrinos}, probe cosmological distances. Therefore, the UHE neutrino flux produced via the GZK effect could still be significant if either: sources beyond ${\sim}100$~Mpc of Earth (roughly, $z\gtrsim 1$) produce a larger proton flux than sources within this volume; or, the luminosity-density of sources peaks at high redshifts. The former possibility is a question of how ``standard'' UHECR sources are (i.e. how similar they are to one another), while the latter is a question of the underlying UHECR source evolution (i.e. how their sources are distributed in the universe \& how their luminosity changes with redshift). For simplicity, many studies assume that sources are standard and any variation is fully captured by their source evolution. The evolution of sources is typically split into three categories: \textit{positive} source evolutions, which assume sources are more numerous/luminous at high redshifts; \textit{negative} source evolutions, which assume source are more luminous/numerous today than they were in the past; and, \textit{uniform} source evolutions, which assume sources are equally numerous/luminous to high redshifts. In all three cases, it is assumed that a cutoff sets in at high redshifts. A number of models for the source evolution exist in the literature, both parametric (see e.g. equation 2 and Fig. 2 of~\cite{Muzio:2019leu}) and observationally-motivated (e.g.~\cite{Robertson:2015uda,Stanev:2008un,Yuksel:2008cu}). The source evolution itself is not currently well-constrained by observations model-independently. This adds significant uncertainty to the level of the UHE neutrino flux since more positive source evolutions will produce larger fluxes than negative source evolutions.

\par
If, in fact, the source evolution is negative the flux of cosmogenic neutrinos is necessarily minimized. However, one final possibility for production of UHE neutrinos remains: neutrinos produced via UHECR interactions inside their sources -- often called \textit{source} or \textit{astrophysical neutrinos}. Source neutrinos can be produced via both photohadronic and hadronic interactions, and their flux is very poorly constrained due to the wide variety of possible conditions which exist inside sources. In particular, these interactions may cause few UHECRs to actually escape sources, meaning that it is possible for UHECR sources to evolve negatively and still produce a large UHE neutrino flux while remaining compatible with UHECR composition measurements. In fact, it is possible that UHECR interactions inside their source environments result in distinct UHECR-bright and UHE neutrino-bright populations, even if the underlying accelerator of UHECRs is the same. 

\section{Model-independent benchmarks}\label{sec:benchmarks}

\subsection{The UHE neutrino floor}

\par
The minimal UHE neutrino flux which is compatible with data would be the cosmogenic flux of neutrinos produced by the UHECRs observed by the Pierre Auger Observatory and the Telescope Array. To understand the order of magnitude level of this \textit{UHE neutrino floor}, one can turn to studies like~\cite{PierreAuger:2022atd} which perform a combined fit to the UHECR spectrum and composition data.\footnote{For a more robust investigation into the UHE neutrino floor see, e.g.,~\cite{Berat:2024rvf}} In~\cite{PierreAuger:2022atd}, a simple model assuming standard sources is employed. Free parameters control the relative proportions of five mass groups which escape these sources following a single-power law spectrum with an exponential cutoff. These mass groups are assumed to share a common spectral index and maximum rigidity\footnote{Rigidity is the energy per charge, $R\equiv E/Z$.} (often referred to as a \textit{Peters cycle}~\citep{Peters:1961mxb}), since sources are expected to accelerate CRs until they can no longer magnetically confine them. Finally, a second CR component, either of Galactic or extragalactic origin, is introduced to fit UHECR data below $10^{18.6}$~eV. These parameters are optimized for a fixed source evolution to give the best-fit to the observed spectrum and composition.

\par
\begin{figure}[htpb!]
    \centering
    \includegraphics[width=0.8\linewidth]{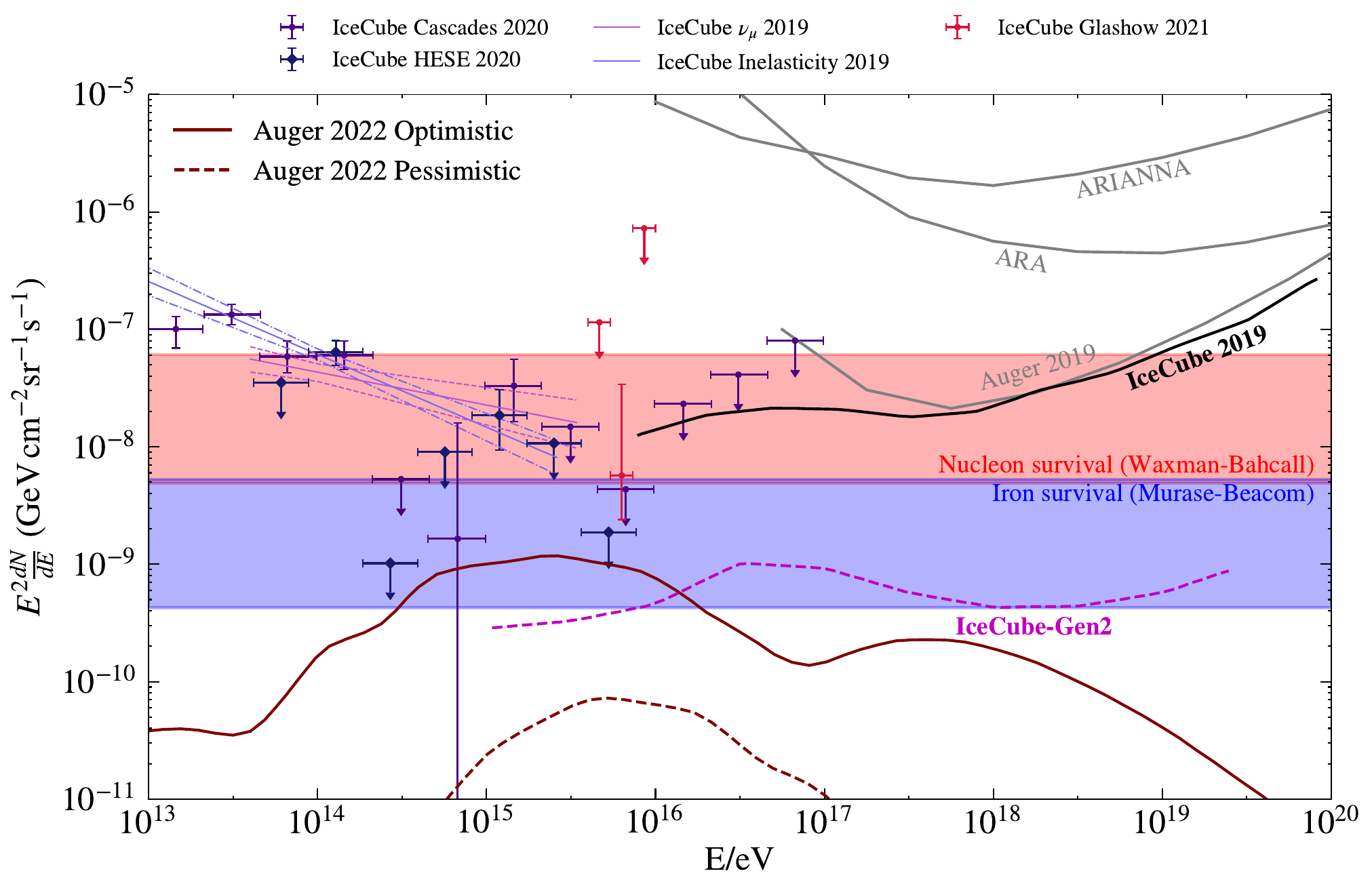}
    \caption{Summary of model-independent benchmarks for the UHE neutrino flux. Data and models are detailed in the text.}
    \label{fig:benchmarks}
\end{figure}

The neutrino floor for a strong negative (pessimistic) and strong positive (optimistic) source evolution are shown in Fig.~\ref{fig:benchmarks}, along with existing data~\citep{IceCube:2018pgc,Stettner:2019tok,IceCube:2020acn,IceCube:2020wum,IceCube:2021rpz} and upper-bounds~\citep{IceCube:2018fhm,PierreAuger:2019ens,ARA:2019wcf,Anker:2019rzo}, as well as, the projected sensitivity of the proposed IceCube-Gen2 experiment~\citep{IceCube-Gen2:2023vtj} for reference. While the uncertainty in the UHE neutrino floor due to source evolution spans more than an order of magnitude, it is clear that current UHECR data is compatible with a very low UHE neutrino flux. This is primarily due to the mixed composition observed above $10^{18.6}$~eV which appears to become heavier at higher energies. Since nuclei are heavier, their energy-per-nucleon (the quantity which primarily governs the energy of secondary neutrinos) remains moderate, $\lesssim 10$~EeV even at high energies. Based on these current estimates of the UHE neutrino floor, the next-generation of UHE neutrino experiments would only be able to detect the falling edge of this spectrum in the $10-100$~PeV range in the most optimistic scenarios. However, robust detection of a population of light nuclei or protons above $10$~EeV by experiments like AugerPrime~\citep{PierreAuger:2016qzd} could significantly increase the level of the neutrino floor at UHEs. 

\subsection{The nucleon survival bound}

A second benchmark for the UHE neutrino flux comes from the observation that in order for UHECRs to reach Earth, they must escape their sources without significant energy losses due to interactions. In this section we closely follow the discussion in~\cite{Anchordoqui:2018qom}. 

\par
The integral energy flux of the observed UHECR spectrum gives a benchmark for the local UHECR energy density

\begin{align}
    \epsilon_\mathrm{CR} = \frac{4\pi}{c} \int EJ(E) dE~.
\end{align}

\noindent
In the $10-1000$~EeV range this corresponds to roughly $\epsilon_\mathrm{CR} \simeq 10^{-19}$~TeV/cm$^3$ and would require a population emitting over the Hubble time with a CR luminosity-density of $\dot{\epsilon}_\mathrm{CR} \simeq 5\times 10^{44}$~erg/Mpc$^3$/yr. This translates into an energy-dependent CR energy-generation rate of 

\begin{align} 
    E^2\dot{n}_\mathrm{CR}(E) = \frac{\dot{\epsilon}_\mathrm{CR}}{\ln(E_\mathrm{max}/E_\mathrm{min})} \simeq 10^{44}\text{ erg/Mpc$^3$/yr} 
\end{align}

\noindent
between $10$ and $100$~EeV. This rate places a rough upper-bound on the UHE neutrino flux since the UHECRs we observe, whose flux corresponds to this CR energy-generation rate, must have few enough neutrino-producing interactions to survive their journey to Earth. 

\par
Such a calculation is relatively simple if one assumes all UHECRs are protons. Of course, we know this is not the case from UHECR composition data, but it is still a useful exercise in terms of benchmarking the UHE neutrino flux. Protons inside a source will primarily undergo photopion production interactions, assuming their interactions are photohadronically-dominated. Let's assume a fraction $f_\pi$ of their energy is lost to pions. In order that a substantial UHECR flux is observed this fraction cannot be too large. Therefore, we require that $f_\pi < 1/2$ --- the \textit{nucleon survival condition} --- so that sources are in the semi-transparent regime. We can write the energy density of photopion produced neutrinos in terms of $f_\pi$ as $\epsilon_\nu(E) \simeq \frac{1}{2} \times \frac{3}{4} f_\pi \xi_z t_\mathrm{H} \frac{1}{1-f_\pi} E^2 \dot{n}_\mathrm{CR}(E)$, where $\xi_z \equiv \int_0^\infty dz \frac{(1+z)^\gamma}{\sqrt{\Omega_M (1+z)^3 + \Omega_\Lambda}} \xi(z)$ is a factor accounting for the UHECR source evolution $\xi(z)$ given a spectral index $E^\gamma$ (where $z$ is the redshift, and $\Omega_M$ and $\Omega_\Lambda$ represent the matter and dark energy density, respectively). For realistic source evolutions $\xi_z\sim \mathcal{O}(0.1-10)$. In the previous expression, $1/2$ corresponds to the fraction of pions which are charged,\footnote{Since most photopion production interactions occur near threshold, pions are produced through $\Delta$-resonance decay. Due to charge conservation, this decay can only produce a single charged pion or a $\pi^0$. Namely, $n+\gamma\rightarrow \Delta^0 \rightarrow n/p + \pi^0/\pi^-$ while $p+\gamma \rightarrow \Delta^+ \rightarrow p/n + \pi^0/\pi^+$. For each of these decays, the charged and neutral pion final states are produced in approximately equal numbers due to isospin symmetry.} $3$ corresponds to the number of neutrinos produced by each charged pion, and $1/4$ is roughly the fraction of the pion energy carried by each neutrino. Finally, $(1-f_\pi)^{-1} E^2 \dot{n}_\mathrm{CR}$ gives the injected CR energy-generation rate. The energy flux of neutrinos $E_\nu^2 \phi(E_\nu) = \frac{c}{4\pi} \epsilon_\nu$ is then saturated when $f_\pi = 1/2$,

\begin{align}\label{eq:nucleon_survival}
    E_\nu^2 \phi(E_\nu) \lesssim 2.3\times 10^{-8} \xi_z\text{ GeV/cm$^2$/s/sr}~.
\end{align}

\noindent
This expression is often called the \textit{nucleon survival bound}, or the \textit{Waxman-Bahcall bound}~\citep{Waxman:1998yy,Bahcall:1999yr}. The corresponding UHE neutrino flux, assuming an $E^{-2}$ spectrum, is shown in Fig.~\ref{fig:benchmarks}. As we will see, this bound provides a rough theoretical estimate of the \textit{UHE neutrino ceiling}.

\subsection{The nucleus survival bound}

\par
\begin{figure}[htpb!]
    \centering
    \includegraphics[width=0.6\linewidth]{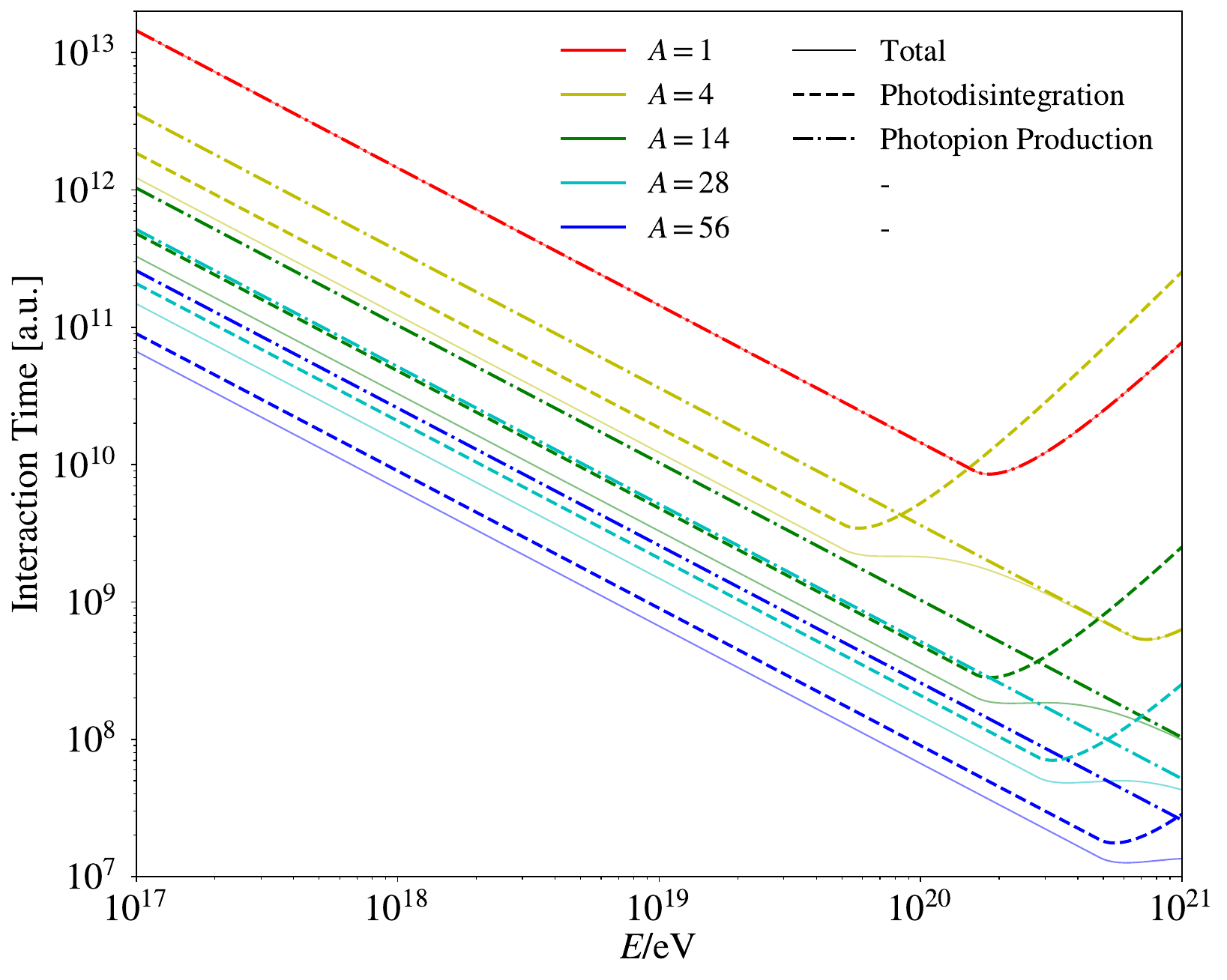}
    \caption{Interaction times (in the narrow-width approximation) for a CR propagating through a photon field with a broken power law spectral density. The ratio of photodisintegration and photopion production interaction times for nuclei are approximately constant up to high energies.}
    \label{fig:PD2PP}
\end{figure}

A similar bound to~\eqref{eq:nucleon_survival} can be obtained for nuclei. As mentioned in Section~\ref{sec:UHECRdata}, nuclei undergoing photohadronic interactions will primarily photodisintegrate rather than photopion produce. Therefore a comparable nucleus survival condition can be written as $f_\mathrm{PD} < 1/2$, where $f_\mathrm{PD}$ is the fraction of the total primary CR energy lost to photodisintegration. These interactions will not lead to a corresponding neutrino flux, but under mild assumptions about the target photon field's spectrum (namely, that it follows a power law) the ratio of photopion production-to-photodisintegration interactions for nuclei is approximately constant to very high energies, as illustrated in Fig.~\ref{fig:PD2PP}. Under a narrow-width approximation, this constant ratio depends only on fundamental quantities, such as the cross section and resonance width. The corresponding energy loss fractions for the two processes are therefore related by $f_\mathrm{PP} \simeq 8.2\times 10^{-2} (A/56)^{-0.21} f_\mathrm{PD}$ for nuclei $A>4$, where we now write $f_\mathrm{PP}$ for the photopion energy loss fraction for nuclei to distinguish from the pure-proton case above. We omit the corresponding expression for $A\leq 4$, which can be derived following~\cite{Murase:2010gj}. 

\par
Noting that the bulk of energy loss is due to photodisintegration but all neutrino production is due to photopion production, we can write the resulting neutrino energy density as $\epsilon_\nu(E) \simeq \frac{1}{2} \times \frac{3}{4} f_\mathrm{PP} \xi_z t_\mathrm{H} \frac{1}{1-f_\mathrm{PD}} E^2 \dot{n}_\mathrm{CR}(E)$. The corresponding neutrino energy flux is then saturated when $f_\mathrm{PD} = 1/2$,

\begin{align}\label{eq:nucleus_survival}
    E_\nu^2 \phi(E_\nu) \lesssim 2\times 10^{-9} \xi_z \left(A/56\right)^{-0.21}\text{ GeV/cm$^2$/s/sr}~.
\end{align}

\noindent
This bound is often called the \textit{nucleus survival bound}, or the  \textit{Murase-Beacom bound}~\citep{Murase:2010gj}, and is plotted for the extreme case of iron nuclei assuming an $E^{-2}$ spectrum in Fig.~\ref{fig:benchmarks}. The resulting benchmark is smaller than that for nucleons, since photopion production is suppressed for nuclei, and decreases for heavier nuclei, since their energy-per-nucleon is smaller for a given energy.

\section{Model-dependent predictions}\label{sec:predictions}

\par
Fundamentally any model which predicts a UHE neutrino flux produced by UHECRs must be compatible with UHECR spectrum \& composition observations. In addition, the resulting secondary fluxes of neutrinos, as well as, gamma-rays\footnote{We note that gamma-rays are produced via neutral pion decay, $\pi^0 \rightarrow \gamma\gamma$, each with energy $E_\gamma = \frac{1}{2} E_\pi$. Therefore, UHECR source models must also be compatible with gamma-ray data \& constraints.} predicted by these models should be compatible with existing bounds and observations on these messengers at all energies. But beyond these basic considerations model-dependent predictions of the UHE neutrino flux span a broad range of assumptions and flux levels --- and many such models exist in the literature. These models, however, can be categorized into two broad classes: top-down and bottom-up UHECR source models.

\par
In order to make specific predictions of the UHECR \& secondary neutrino fluxes, a model must at least account for UHECR interactions during extragalactic propagation. This requires assuming a UHECR flux escaping sources and a source evolution. After these assumptions are made one must solve transport equations to obtain the flux of UHECRs \& neutrinos arriving at Earth. A number of standard software packages (not all of which produce secondary particle predictions) are available for this purpose, including~\cite{AlvesBatista:2022vem,Aloisio:2017iyh,Heinze:2019jou}. 

\begin{figure}[htpb!]
    \centering
    \includegraphics[width=\linewidth]{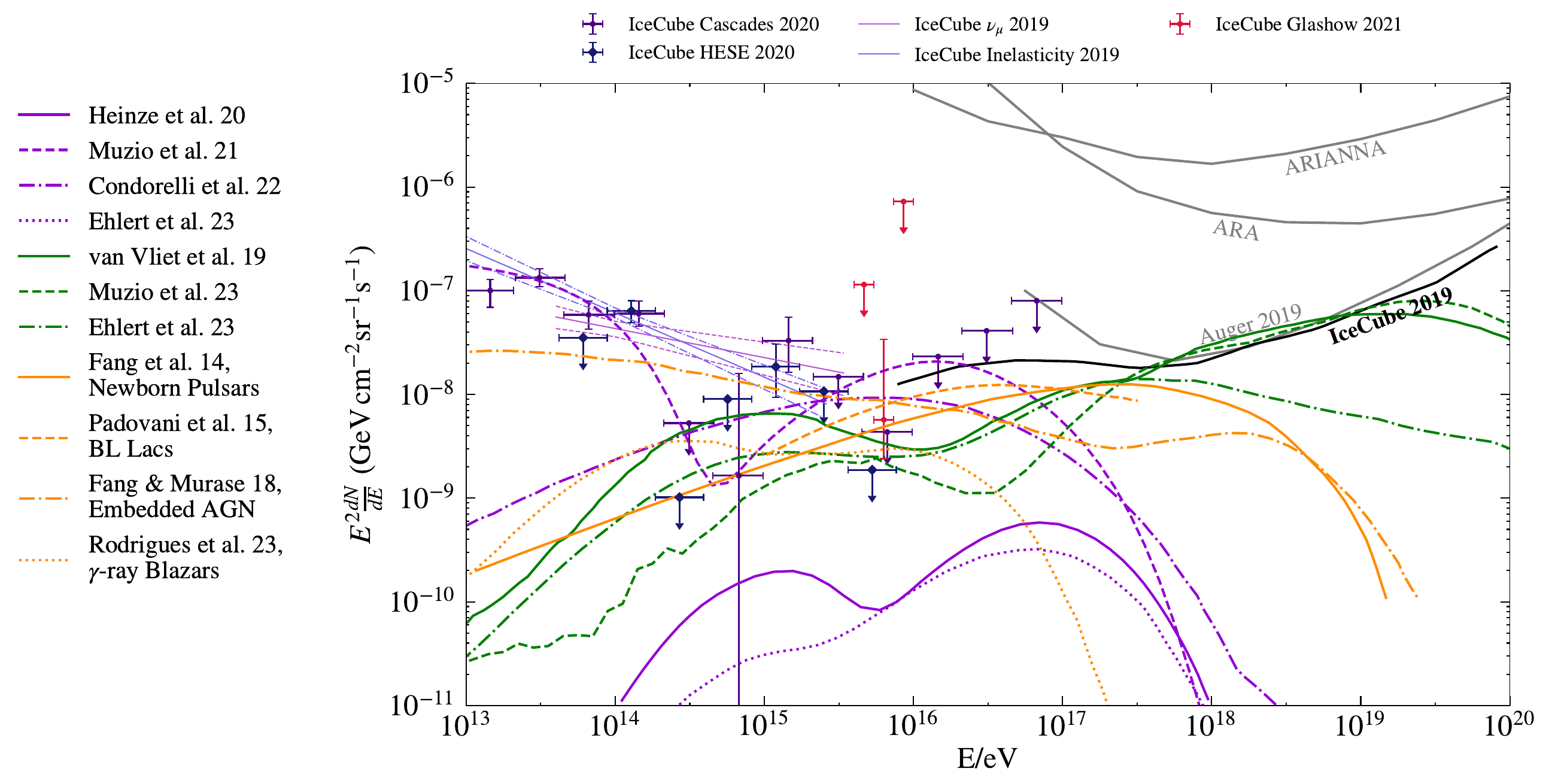}
    \caption{Summary of model-dependent predictions for the UHE neutrino flux. Data and models are detailed in the text.}
    \label{fig:modelfluxes}
\end{figure}

\subsection{Top-down UHECR source models}

Top-down models take a more phenomenological, data-driven approach to model UHECR data. Very broadly, a model is developed which captures the essential physical processes but which has a number of unknown free parameters. These parameters are then tuned to provide the best-fit to UHECR spectrum \& composition data, ideally accounting for neutrino and gamma-ray constraints. These models generally provide a more accurate description of the data, but their best-fit parameters may not be immediately connected to any particular astrophysical source or be in a physical part of the parameter space. Simply put, these models determine the properties sources would need to achieve the best-fit to the UHECR data. Generally, since these models are mainly concerned with explaining the UHECR data they do not necessarily explore the full range of UHE neutrino fluxes which are compatible with that data. Figure~\ref{fig:modelfluxes} shows (purple lines) some examples~\citep{Heinze:2020zqb,Muzio:2021zud,Condorelli:2022vfa,Ehlert:2023btz} of UHE neutrino fluxes predicted by top-down UHECR source models from the literature.

\subsubsection{UHECR-compatible trans-GZK models}

\par
UHECR-compatible trans-GZK models attempt to fully explore the range of possible UHE neutrino fluxes compatible with UHECR data --- in other words, they seek to answer the question ``how large \textit{could} the UHE neutrino flux be?'' In this sense, they represent a specific realization of the UHE neutrino flux ceiling, reflecting both constraints from observation and what is possible theoretically. Typically, such models assume two separate populations of UHECR sources exist: a population producing a mixed composition flux with a maximum rigidity $<10$~EV which describes the bulk of the UHECR data; and a population producing a (usually pure-proton) flux with a maximum rigidity $>10$~EV and up to ${\sim}1000$~EV which produces the bulk of UHE neutrinos. The main UHECR population's parameters are fit as described above, though they may be mildly affected by the second UHECR population. The second UHECR population's parameters are often tuned to maximize the UHE neutrino flux while remaining compatible with UHECR composition data and neutrino \& gamma-ray constraints. Figure~\ref{fig:modelfluxes} shows (green lines) some examples~\citep{vanVliet:2019nse,Anker:2020lre,Muzio:2023skc,Ehlert:2023btz} of UHE neutrino fluxes predicted by UHECR-compatible trans-GZK models from the literature.

\subsection{Bottom-up UHECR source models}

\par
Bottom-up models take a more theoretical, observation-driven approach to model UHECR data. Generally, bottom-up models hypothesize a particular source type (e.g. AGNs, GRBs, etc.) and what the relevant acceleration and energy loss processes are. These hypotheses strongly constrain the parameters of the model including the composition of injected cosmic rays, the maximum rigidity, and the source evolution, among others. The resulting model may require detailed simulation of CR propagation within the environment host to the accelerator to be fully evaluated. Once this is done, the observed UHECR spectrum \& composition and neutrino \& gamma-ray fluxes are predicted and compared to data. These models generally provide a worse description of UHECR data but can be directly tied to the plausibility of particular source types and properties explaining this data. To put it simply, these models demonstrate the ability of known astrophysical sources, as we currently understand them, to explain UHECR observations. In some cases, these models predict a larger UHE neutrino flux than typical top-down models, especially at high energies, since they do not directly confine themselves to the properties sufficient to describe UHECR observations. Figure~\ref{fig:modelfluxes} shows (orange lines) some examples~\citep{Fang:2013vla,Padovani:2015mba,Fang:2017zjf,Rodrigues:2023vbv} of UHE neutrino fluxes predicted by bottom-up UHECR source models from the literature.

\section{Summary: The UHE neutrino flux zoo}\label{sec:summary}

\par
Ultrahigh energy neutrinos have the potential to provide unparalleled probes into the highest energy astrophysical environments in the Universe, as well as, into particle physics beyond LHC energies and beyond the Standard Model. Today robust observations of the UHECR spectrum \& composition have provided strong evidence that UHE neutrinos exist --- but the exact level of their flux remains an open question. 

\par
\begin{figure}[htpb!]
    \centering
    \includegraphics[width=\linewidth]{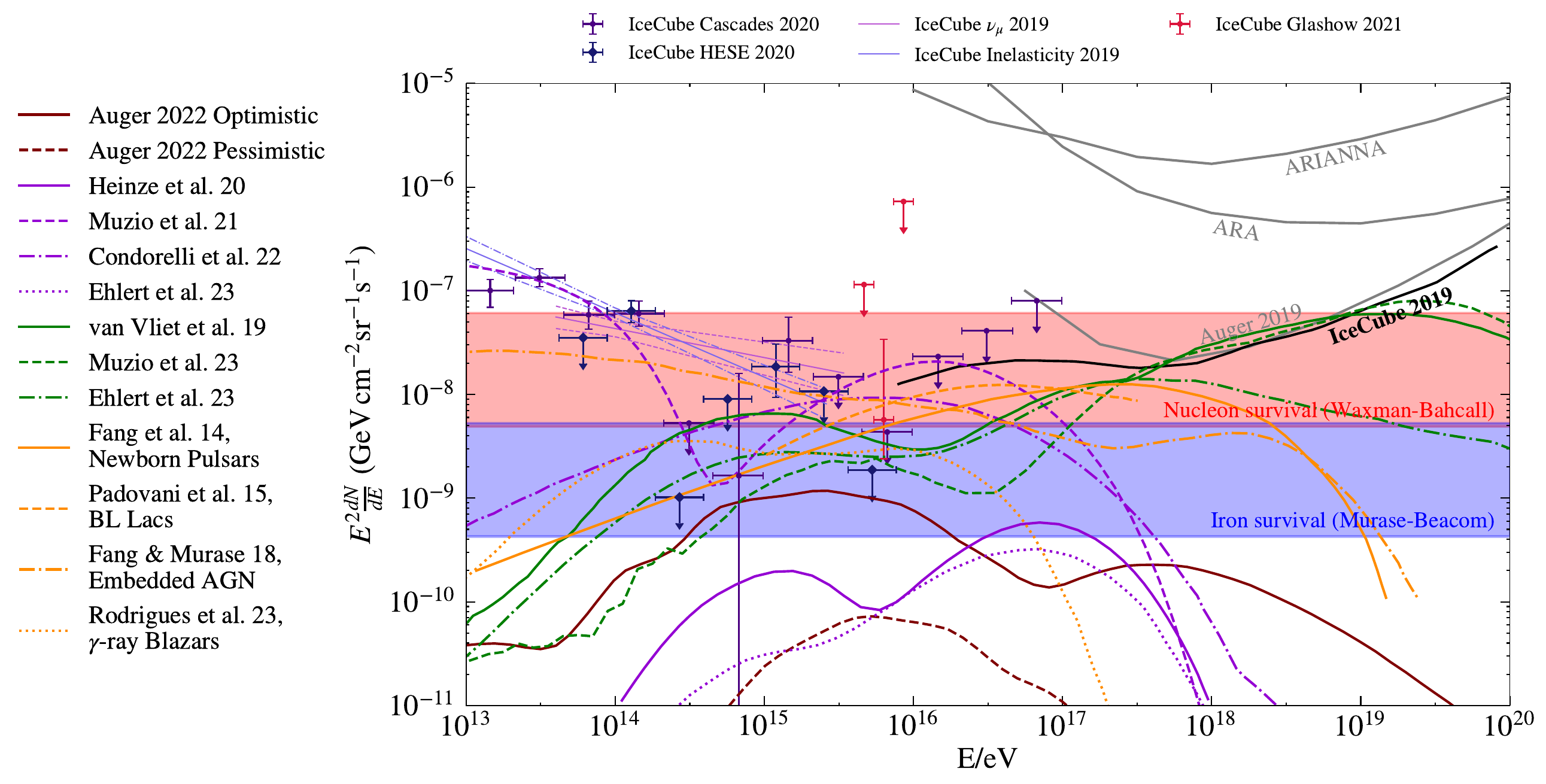}
    \caption{The UHE neutrino flux zoo: a summary of the model-independent benchmarks and model predictions for the UHE neutrino flux discussed in this \paper{}. This summary is not exhaustive.}
    \label{fig:uheNuZoo}
\end{figure}

Throughout this review, we have discussed broad approaches to characterizing the UHE neutrino flux and the data which constrains it. Model-independent approaches provide rough benchmarks of both the UHE neutrino floor and its ceiling. But between these boundaries, a diverse set of model-dependent predictions exist. These models, for which we have included a woefully inexhaustive selection of examples, span a broad range of assumptions and predictions -- which collectively make up a kind of UHE neutrino flux zoo, as illustrated in Fig.~\ref{fig:uheNuZoo}. 

\par
Today, no one of these models is significantly better supported by observational data than the others. Only ambitious and innovative experiments will discover which of these represents the true landscape of the UHE neutrino frontier. Luckily a number of next-generation UHE neutrino experiments, using a diverse set of techniques, have been proposed. A discussion of these next-generation experiments can be found in~\cite{Ackermann:2022rqc}. To be successful, they will require the next-generation of astroparticle physicists to bring to fruition the enormous exposures needed to illuminate the UHE neutrino landscape -- and with it, perhaps, the true origin of UHECRs.

\bibliography{main}

\begin{thebibliography}{10}
\providecommand{\url}[1]{{#1}}
\providecommand{\urlprefix}{URL }
\providecommand{\doi}[1]{\url{https://doi.org/#1}}
\bibcommenthead

\bibitem{IceCube:2013low}
M.G. Aartsen, et~al., {Evidence for High-Energy Extraterrestrial Neutrinos at the IceCube Detector}.
\newblock Science \textbf{342}, 1242856 (2013).
\newblock \doi{10.1126/science.1242856}.
\newblock {\href{https://arxiv.org/abs/1311.5238}{{arXiv:1311.5238}}} {[astro-ph.HE]}

\bibitem{Naab:2023xcz}
R.~Naab, E.~Ganster, Z.~Zhang, \emph{{Measurement of the astrophysical diffuse neutrino flux in a combined fit of IceCube's high energy neutrino data}}, in \emph{{38th International Cosmic Ray Conference}} (2023)

\bibitem{Silva:2023wol}
M.~Silva, S.~Mancina, J.~Osborn, {Measurement of the Cosmic Neutrino Flux from the Southern Sky using 10 years of IceCube Starting Track Events}.
\newblock PoS \textbf{ICRC2023}, 1008 (2023).
\newblock \doi{10.22323/1.444.1008}.
\newblock {\href{https://arxiv.org/abs/2308.04582}{{arXiv:2308.04582}}} {[astro-ph.HE]}

\bibitem{IceCube:2022der}
R.~Abbasi, et~al., {Evidence for neutrino emission from the nearby active galaxy NGC 1068}.
\newblock Science \textbf{378}(6619), 538--543 (2022).
\newblock \doi{10.1126/science.abg3395}.
\newblock {\href{https://arxiv.org/abs/2211.09972}{{arXiv:2211.09972}}} {[astro-ph.HE]}

\bibitem{IceCube:2018cha}
M.G. Aartsen, et~al., {Neutrino emission from the direction of the blazar TXS 0506+056 prior to the IceCube-170922A alert}.
\newblock Science \textbf{361}(6398), 147--151 (2018).
\newblock \doi{10.1126/science.aat2890}.
\newblock {\href{https://arxiv.org/abs/1807.08794}{{arXiv:1807.08794}}} {[astro-ph.HE]}

\bibitem{IceCube:2018dnn}
M.G. Aartsen, et~al., {Multimessenger observations of a flaring blazar coincident with high-energy neutrino IceCube-170922A}.
\newblock Science \textbf{361}(6398), eaat1378 (2018).
\newblock \doi{10.1126/science.aat1378}.
\newblock {\href{https://arxiv.org/abs/1807.08816}{{arXiv:1807.08816}}} {[astro-ph.HE]}

\bibitem{IceCube:2023ame}
R.~Abbasi, et~al., {Observation of high-energy neutrinos from the Galactic plane}.
\newblock Science \textbf{380}(6652), adc9818 (2023).
\newblock \doi{10.1126/science.adc9818}.
\newblock {\href{https://arxiv.org/abs/2307.04427}{{arXiv:2307.04427}}} {[astro-ph.HE]}

\bibitem{IceCube:2021rpz}
M.G. Aartsen, et~al., {Detection of a particle shower at the Glashow resonance with IceCube}.
\newblock Nature \textbf{591}(7849), 220--224 (2021).
\newblock \doi{10.1038/s41586-021-03256-1}.
\newblock [Erratum: Nature 592, E11 (2021)].
\newblock {\href{https://arxiv.org/abs/2110.15051}{{arXiv:2110.15051}}} {[hep-ex]}

\bibitem{IceCube:2024nhk}
R.~Abbasi, et~al., {Observation of Seven Astrophysical Tau Neutrino Candidates with IceCube}.
\newblock Phys. Rev. Lett. \textbf{132}(15), 151001 (2024).
\newblock \doi{10.1103/PhysRevLett.132.151001}.
\newblock {\href{https://arxiv.org/abs/2403.02516}{{arXiv:2403.02516}}} {[astro-ph.HE]}

\bibitem{Ackermann:2022rqc}
M.~Ackermann, et~al., {High-energy and ultra-high-energy neutrinos: A Snowmass white paper}.
\newblock JHEAp \textbf{36}, 55--110 (2022).
\newblock \doi{10.1016/j.jheap.2022.08.001}.
\newblock {\href{https://arxiv.org/abs/2203.08096}{{arXiv:2203.08096}}} {[hep-ph]}

\bibitem{Linsley:1963km}
J.~Linsley, {Evidence for a primary cosmic-ray particle with energy 10**20-eV}.
\newblock Phys. Rev. Lett. \textbf{10}, 146--148 (1963).
\newblock \doi{10.1103/PhysRevLett.10.146}

\bibitem{Anchordoqui:2018qom}
L.A. Anchordoqui, {Ultra-High-Energy Cosmic Rays}.
\newblock Phys. Rept. \textbf{801}, 1--93 (2019).
\newblock \doi{10.1016/j.physrep.2019.01.002}.
\newblock {\href{https://arxiv.org/abs/1807.09645}{{arXiv:1807.09645}}} {[astro-ph.HE]}

\bibitem{pdg24}
S.~Navas, et~al., {Review of Cosmic Rays}  (2024).
\newblock \urlprefix\url{https://pdg.lbl.gov/2024/reviews/rpp2024-rev-cosmic-rays.pdf}.
\newblock To be published in Phys. Rev. D 110, 030001 (2024)

\bibitem{PierreAuger:2023wti}
D.R. Bergman, et~al., {Measurement of UHECR energy spectrum with the Pierre Auger Observatory and the Telescope Array}.
\newblock PoS \textbf{ICRC2023}, 406 (2024).
\newblock \doi{10.22323/1.444.0406}

\bibitem{Greisen:1966jv}
K.~Greisen, {End to the cosmic ray spectrum?}
\newblock Phys. Rev. Lett. \textbf{16}, 748--750 (1966).
\newblock \doi{10.1103/PhysRevLett.16.748}

\bibitem{Zatsepin:1966jv}
G.T. Zatsepin, V.A. Kuzmin, {Upper limit of the spectrum of cosmic rays}.
\newblock JETP Lett. \textbf{4}, 78--80 (1966)

\bibitem{PierreAuger:2024flk}
A.~Abdul~Halim, et~al., {Inference of the Mass Composition of Cosmic Rays with energies from $\mathbf{10^{18.5}}$ to $\mathbf{10^{20}}$ eV using the Pierre Auger Observatory and Deep Learning}  (2024).
\newblock {\href{https://arxiv.org/abs/2406.06315}{{arXiv:2406.06315}}} {[astro-ph.HE]}

\bibitem{PierreAuger:2024nzw}
A.~Abdul~Halim, et~al., {Measurement of the Depth of Maximum of Air-Shower Profiles with energies between $\mathbf{10^{18.5}}$ and $\mathbf{10^{20}}$ eV using the Surface Detector of the Pierre Auger Observatory and Deep Learning}  (2024).
\newblock {\href{https://arxiv.org/abs/2406.06319}{{arXiv:2406.06319}}} {[astro-ph.HE]}

\bibitem{Ostapchenko:2010vb}
S.~Ostapchenko, {Monte Carlo treatment of hadronic interactions in enhanced Pomeron scheme: I. QGSJET-II model}.
\newblock Phys. Rev. D \textbf{83}, 014018 (2011).
\newblock \doi{10.1103/PhysRevD.83.014018}.
\newblock {\href{https://arxiv.org/abs/1010.1869}{{arXiv:1010.1869}}} {[hep-ph]}

\bibitem{Pierog:2013ria}
T.~Pierog, I.~Karpenko, J.M. Katzy, E.~Yatsenko, K.~Werner, {EPOS LHC: Test of collective hadronization with data measured at the CERN Large Hadron Collider}.
\newblock Phys. Rev. C \textbf{92}(3), 034906 (2015).
\newblock \doi{10.1103/PhysRevC.92.034906}.
\newblock {\href{https://arxiv.org/abs/1306.0121}{{arXiv:1306.0121}}} {[hep-ph]}

\bibitem{Riehn:2019jet}
F.~Riehn, R.~Engel, A.~Fedynitch, T.K. Gaisser, T.~Stanev, {Hadronic interaction model Sibyll 2.3d and extensive air showers}.
\newblock Phys. Rev. D \textbf{102}(6), 063002 (2020).
\newblock \doi{10.1103/PhysRevD.102.063002}.
\newblock {\href{https://arxiv.org/abs/1912.03300}{{arXiv:1912.03300}}} {[hep-ph]}

\bibitem{PierreAuger:2023yym}
A.~Abdul~Halim, et~al., {Depth of maximum of air-shower profiles: testing the compatibility of the measurements at the Pierre Auger Observatory and the Telescope Array}.
\newblock PoS \textbf{ICRC2023}, 249 (2023).
\newblock \doi{10.22323/1.444.0249}

\bibitem{Muzio:2019leu}
M.S. Muzio, M.~Unger, G.R. Farrar, {Progress towards characterizing ultrahigh energy cosmic ray sources}.
\newblock Phys. Rev. D \textbf{100}(10), 103008 (2019).
\newblock \doi{10.1103/PhysRevD.100.103008}.
\newblock {\href{https://arxiv.org/abs/1906.06233}{{arXiv:1906.06233}}} {[astro-ph.HE]}

\bibitem{Muzio:2023skc}
M.S. Muzio, M.~Unger, S.~Wissel, {Prospects for joint cosmic ray and neutrino constraints on the evolution of trans-Greisen-Zatsepin-Kuzmin proton sources}.
\newblock Phys. Rev. D \textbf{107}(10), 103030 (2023).
\newblock \doi{10.1103/PhysRevD.107.103030}.
\newblock {\href{https://arxiv.org/abs/2303.04170}{{arXiv:2303.04170}}} {[astro-ph.HE]}

\bibitem{Robertson:2015uda}
B.E. Robertson, R.S. Ellis, S.R. Furlanetto, J.S. Dunlop, {Cosmic Reionization and Early Star-forming Galaxies: a Joint Analysis of new Constraints From Planck and the Hubble Space Telescope}.
\newblock Astrophys. J. Lett. \textbf{802}(2), L19 (2015).
\newblock \doi{10.1088/2041-8205/802/2/L19}.
\newblock {\href{https://arxiv.org/abs/1502.02024}{{arXiv:1502.02024}}} {[astro-ph.CO]}

\bibitem{Stanev:2008un}
T.~Stanev, {Ultra high energy cosmic rays and neutrinos after Auger}  (2008).
\newblock {\href{https://arxiv.org/abs/0808.1045}{{arXiv:0808.1045}}} {[astro-ph]}

\bibitem{Yuksel:2008cu}
H.~Yuksel, M.D. Kistler, J.F. Beacom, A.M. Hopkins, {Revealing the High-Redshift Star Formation Rate with Gamma-Ray Bursts}.
\newblock Astrophys. J. Lett. \textbf{683}, L5--L8 (2008).
\newblock \doi{10.1086/591449}.
\newblock {\href{https://arxiv.org/abs/0804.4008}{{arXiv:0804.4008}}} {[astro-ph]}

\bibitem{PierreAuger:2022atd}
A.A. Halim, et~al., {Constraining the sources of ultra-high-energy cosmic rays across and above the ankle with the spectrum and composition data measured at the Pierre Auger Observatory}.
\newblock JCAP \textbf{05}, 024 (2023).
\newblock \doi{10.1088/1475-7516/2023/05/024}.
\newblock {\href{https://arxiv.org/abs/2211.02857}{{arXiv:2211.02857}}} {[astro-ph.HE]}

\bibitem{Berat:2024rvf}
C.~Berat, A.~Condorelli, O.~Deligny, F.~Montanet, Z.~Torres, {Floor of Cosmogenic Neutrino Fluxes above 10$^{17}$ eV}.
\newblock Astrophys. J. \textbf{966}(2), 186 (2024).
\newblock \doi{10.3847/1538-4357/ad372a}.
\newblock {\href{https://arxiv.org/abs/2402.04759}{{arXiv:2402.04759}}} {[astro-ph.HE]}

\bibitem{Peters:1961mxb}
B.~Peters, {Primary cosmic radiation and extensive air showers}.
\newblock Nuovo Cim. \textbf{22}(4), 800--819 (1961).
\newblock \doi{10.1007/bf02783106}

\bibitem{IceCube:2018pgc}
M.G. Aartsen, et~al., {Measurements using the inelasticity distribution of multi-TeV neutrino interactions in IceCube}.
\newblock Phys. Rev. D \textbf{99}(3), 032004 (2019).
\newblock \doi{10.1103/PhysRevD.99.032004}.
\newblock {\href{https://arxiv.org/abs/1808.07629}{{arXiv:1808.07629}}} {[hep-ex]}

\bibitem{Stettner:2019tok}
J.~Stettner, {Measurement of the Diffuse Astrophysical Muon-Neutrino Spectrum with Ten Years of IceCube Data}.
\newblock PoS \textbf{ICRC2019}, 1017 (2020).
\newblock \doi{10.22323/1.358.1017}.
\newblock {\href{https://arxiv.org/abs/1908.09551}{{arXiv:1908.09551}}} {[astro-ph.HE]}

\bibitem{IceCube:2020acn}
M.G. Aartsen, et~al., {Characteristics of the diffuse astrophysical electron and tau neutrino flux with six years of IceCube high energy cascade data}.
\newblock Phys. Rev. Lett. \textbf{125}(12), 121104 (2020).
\newblock \doi{10.1103/PhysRevLett.125.121104}.
\newblock {\href{https://arxiv.org/abs/2001.09520}{{arXiv:2001.09520}}} {[astro-ph.HE]}

\bibitem{IceCube:2020wum}
R.~Abbasi, et~al., {The IceCube high-energy starting event sample: Description and flux characterization with 7.5 years of data}.
\newblock Phys. Rev. D \textbf{104}, 022002 (2021).
\newblock \doi{10.1103/PhysRevD.104.022002}.
\newblock {\href{https://arxiv.org/abs/2011.03545}{{arXiv:2011.03545}}} {[astro-ph.HE]}

\bibitem{IceCube:2018fhm}
M.G. Aartsen, et~al., {Differential limit on the extremely-high-energy cosmic neutrino flux in the presence of astrophysical background from nine years of IceCube data}.
\newblock Phys. Rev. D \textbf{98}(6), 062003 (2018).
\newblock \doi{10.1103/PhysRevD.98.062003}.
\newblock {\href{https://arxiv.org/abs/1807.01820}{{arXiv:1807.01820}}} {[astro-ph.HE]}

\bibitem{PierreAuger:2019ens}
A.~Aab, et~al., {Probing the origin of ultra-high-energy cosmic rays with neutrinos in the EeV energy range using the Pierre Auger Observatory}.
\newblock JCAP \textbf{10}, 022 (2019).
\newblock \doi{10.1088/1475-7516/2019/10/022}.
\newblock {\href{https://arxiv.org/abs/1906.07422}{{arXiv:1906.07422}}} {[astro-ph.HE]}

\bibitem{ARA:2019wcf}
P.~Allison, et~al., {Constraints on the diffuse flux of ultrahigh energy neutrinos from four years of Askaryan Radio Array data in two stations}.
\newblock Phys. Rev. D \textbf{102}(4), 043021 (2020).
\newblock \doi{10.1103/PhysRevD.102.043021}.
\newblock {\href{https://arxiv.org/abs/1912.00987}{{arXiv:1912.00987}}} {[astro-ph.HE]}

\bibitem{Anker:2019rzo}
A.~Anker, et~al., {A search for cosmogenic neutrinos with the ARIANNA test bed using 4.5 years of data}.
\newblock JCAP \textbf{03}, 053 (2020).
\newblock \doi{10.1088/1475-7516/2020/03/053}.
\newblock {\href{https://arxiv.org/abs/1909.00840}{{arXiv:1909.00840}}} {[astro-ph.IM]}

\bibitem{IceCube-Gen2:2023vtj}
R.~Abbasi, et~al., {The next generation neutrino telescope: IceCube-Gen2}.
\newblock PoS \textbf{ICRC2023}, 994 (2023).
\newblock \doi{10.22323/1.444.0994}.
\newblock {\href{https://arxiv.org/abs/2308.09427}{{arXiv:2308.09427}}} {[astro-ph.HE]}

\bibitem{PierreAuger:2016qzd}
A.~Aab, et~al., {The Pierre Auger Observatory Upgrade - Preliminary Design Report}  (2016).
\newblock {\href{https://arxiv.org/abs/1604.03637}{{arXiv:1604.03637}}} {[astro-ph.IM]}

\bibitem{Waxman:1998yy}
E.~Waxman, J.N. Bahcall, {High-energy neutrinos from astrophysical sources: An Upper bound}.
\newblock Phys. Rev. D \textbf{59}, 023002 (1999).
\newblock \doi{10.1103/PhysRevD.59.023002}.
\newblock {\href{https://arxiv.org/abs/hep-ph/9807282}{{arXiv:hep-ph/9807282}}}

\bibitem{Bahcall:1999yr}
J.N. Bahcall, E.~Waxman, {High-energy astrophysical neutrinos: The Upper bound is robust}.
\newblock Phys. Rev. D \textbf{64}, 023002 (2001).
\newblock \doi{10.1103/PhysRevD.64.023002}.
\newblock {\href{https://arxiv.org/abs/hep-ph/9902383}{{arXiv:hep-ph/9902383}}}

\bibitem{Murase:2010gj}
K.~Murase, J.F. Beacom, {Neutrino Background Flux from Sources of Ultrahigh-Energy Cosmic-Ray Nuclei}.
\newblock Phys. Rev. D \textbf{81}, 123001 (2010).
\newblock \doi{10.1103/PhysRevD.81.123001}.
\newblock {\href{https://arxiv.org/abs/1003.4959}{{arXiv:1003.4959}}} {[astro-ph.HE]}

\bibitem{AlvesBatista:2022vem}
R.~Alves~Batista, et~al., {CRPropa 3.2 \textemdash{} an advanced framework for high-energy particle propagation in extragalactic and galactic spaces}.
\newblock JCAP \textbf{09}, 035 (2022).
\newblock \doi{10.1088/1475-7516/2022/09/035}.
\newblock {\href{https://arxiv.org/abs/2208.00107}{{arXiv:2208.00107}}} {[astro-ph.HE]}

\bibitem{Aloisio:2017iyh}
R.~Aloisio, D.~Boncioli, A.~Di~Matteo, A.F. Grillo, S.~Petrera, F.~Salamida, {SimProp v2r4: Monte Carlo simulation code for UHECR propagation}.
\newblock JCAP \textbf{11}, 009 (2017).
\newblock \doi{10.1088/1475-7516/2017/11/009}.
\newblock {\href{https://arxiv.org/abs/1705.03729}{{arXiv:1705.03729}}} {[astro-ph.HE]}

\bibitem{Heinze:2019jou}
J.~Heinze, A.~Fedynitch, D.~Boncioli, W.~Winter, {A new view on Auger data and cosmogenic neutrinos in light of different nuclear disintegration and air-shower models}.
\newblock Astrophys. J. \textbf{873}(1), 88 (2019).
\newblock \doi{10.3847/1538-4357/ab05ce}.
\newblock {\href{https://arxiv.org/abs/1901.03338}{{arXiv:1901.03338}}} {[astro-ph.HE]}

\bibitem{Heinze:2020zqb}
J.~Heinze, D.~Biehl, A.~Fedynitch, D.~Boncioli, A.~Rudolph, W.~Winter, {Systematic parameter space study for the UHECR origin from GRBs in models with multiple internal shocks}.
\newblock Mon. Not. Roy. Astron. Soc. \textbf{498}(4), 5990--6004 (2020).
\newblock \doi{10.1093/mnras/staa2751}.
\newblock {\href{https://arxiv.org/abs/2006.14301}{{arXiv:2006.14301}}} {[astro-ph.HE]}

\bibitem{Muzio:2021zud}
M.S. Muzio, G.R. Farrar, M.~Unger, {Probing the environments surrounding ultrahigh energy cosmic ray accelerators and their implications for astrophysical neutrinos}.
\newblock Phys. Rev. D \textbf{105}(2), 023022 (2022).
\newblock \doi{10.1103/PhysRevD.105.023022}.
\newblock {\href{https://arxiv.org/abs/2108.05512}{{arXiv:2108.05512}}} {[astro-ph.HE]}

\bibitem{Condorelli:2022vfa}
A.~Condorelli, D.~Boncioli, E.~Peretti, S.~Petrera, {Testing hadronic and photohadronic interactions as responsible for ultrahigh energy cosmic rays and neutrino fluxes from starburst galaxies}.
\newblock Phys. Rev. D \textbf{107}(8), 083009 (2023).
\newblock \doi{10.1103/PhysRevD.107.083009}.
\newblock {\href{https://arxiv.org/abs/2209.08593}{{arXiv:2209.08593}}} {[astro-ph.HE]}

\bibitem{Ehlert:2023btz}
D.~Ehlert, A.~van Vliet, F.~Oikonomou, W.~Winter, {Constraints on the proton fraction of cosmic rays at the highest energies and the consequences for cosmogenic neutrinos and photons}.
\newblock JCAP \textbf{02}, 022 (2024).
\newblock \doi{10.1088/1475-7516/2024/02/022}.
\newblock {\href{https://arxiv.org/abs/2304.07321}{{arXiv:2304.07321}}} {[astro-ph.HE]}

\bibitem{vanVliet:2019nse}
A.~van Vliet, R.~Alves~Batista, J.R. H\"orandel, {Determining the fraction of cosmic-ray protons at ultrahigh energies with cosmogenic neutrinos}.
\newblock Phys. Rev. D \textbf{100}(2), 021302 (2019).
\newblock \doi{10.1103/PhysRevD.100.021302}.
\newblock {\href{https://arxiv.org/abs/1901.01899}{{arXiv:1901.01899}}} {[astro-ph.HE]}

\bibitem{Anker:2020lre}
A.~Anker, et~al., {White Paper: ARIANNA-200 high energy neutrino telescope}  (2020).
\newblock {\href{https://arxiv.org/abs/2004.09841}{{arXiv:2004.09841}}} {[astro-ph.IM]}

\bibitem{Fang:2013vla}
K.~Fang, K.~Kotera, K.~Murase, A.V. Olinto, {Testing the Newborn Pulsar Origin of Ultrahigh Energy Cosmic Rays with EeV Neutrinos}.
\newblock Phys. Rev. D \textbf{90}(10), 103005 (2014).
\newblock \doi{10.1103/PhysRevD.90.103005}.
\newblock [Erratum: Phys.Rev.D 92, 129901 (2015)].
\newblock {\href{https://arxiv.org/abs/1311.2044}{{arXiv:1311.2044}}} {[astro-ph.HE]}

\bibitem{Padovani:2015mba}
P.~Padovani, M.~Petropoulou, P.~Giommi, E.~Resconi, {A simplified view of blazars: the neutrino background}.
\newblock Mon. Not. Roy. Astron. Soc. \textbf{452}(2), 1877--1887 (2015).
\newblock \doi{10.1093/mnras/stv1467}.
\newblock {\href{https://arxiv.org/abs/1506.09135}{{arXiv:1506.09135}}} {[astro-ph.HE]}

\bibitem{Fang:2017zjf}
K.~Fang, K.~Murase, {Linking High-Energy Cosmic Particles by Black Hole Jets Embedded in Large-Scale Structures}.
\newblock Nature Phys. \textbf{14}(4), 396--398 (2018).
\newblock \doi{10.1038/s41567-017-0025-4}.
\newblock {\href{https://arxiv.org/abs/1704.00015}{{arXiv:1704.00015}}} {[astro-ph.HE]}

\bibitem{Rodrigues:2023vbv}
X.~Rodrigues, V.S. Paliya, S.~Garrappa, A.~Omeliukh, A.~Franckowiak, W.~Winter, {Leptohadronic multi-messenger modeling of 324 gamma-ray blazars}.
\newblock Astron. Astrophys. \textbf{681}, A119 (2024).
\newblock \doi{10.1051/0004-6361/202347540}.
\newblock {\href{https://arxiv.org/abs/2307.13024}{{arXiv:2307.13024}}} {[astro-ph.HE]}

\end{thebibliography}

\end{document}